\documentclass[twocolumn]{aastex6}

\usepackage{hyperref}
\usepackage{graphicx}
\usepackage{natbib}
\usepackage{amsmath,amsthm,amssymb}
\usepackage{url}

\def\WISE{\textit{WISE}}

\protected\def\Lsed{\ifmmode \,\mathcal{L}_{\mathrm{SED}}\else $\mathcal{L}_{\mathrm{SED}}$\fi}

\protected\def\picometer{\ifmmode \,\operatorname{pm}\else $\operatorname{pm}$\fi}
\protected\def\nm{\ifmmode \,\operatorname{nm}\else $\operatorname{nm}$\fi}
\protected\def\micron{\ifmmode \,\operatorname{\mu m}\else $\operatorname{\mu m}$\fi}
\protected\def\mm{\ifmmode \,\operatorname{mm}\else $\operatorname{mm}$\fi}
\protected\def\meter{\ifmmode \,\operatorname{m}\else $\operatorname{m}$\fi}
\protected\def\km{\ifmmode \,\operatorname{km}\else $\operatorname{km}$\fi}
\protected\def\au{\ifmmode \,\operatorname{AU}\else $\operatorname{AU}$\fi}
\protected\def\pc{\ifmmode \,\operatorname{pc}\else $\operatorname{pc}$\fi}
\protected\def\kpc{\ifmmode \,\operatorname{kpc}\else $\operatorname{kpc}$\fi}
\protected\def\Mpc{\ifmmode \,\operatorname{Mpc}\else $\operatorname{Mpc}$\fi}
\protected\def\rsun{\ifmmode \,\operatorname{R_\odot}\else $\operatorname{R_\odot}$\fi}
\protected\def\Rsun{\ifmmode \,\operatorname{R_\odot}\else $\operatorname{R_\odot}$\fi}

\protected\def\second{\ifmmode \,\operatorname{sec}\else $\operatorname{sec}$\fi}
\protected\def\yr{\ifmmode \,\operatorname{yr}\else $\operatorname{yr}$\fi}
\protected\def\Gyr{\ifmmode \,\operatorname{Gyr}\else $\operatorname{Gyr}$\fi}

\protected\def\eV{\ifmmode \,\operatorname{eV}\else $\operatorname{eV}$\fi}
\protected\def\keV{\ifmmode \,\operatorname{keV}\else $\operatorname{keV}$\fi}
\protected\def\MeV{\ifmmode \,\operatorname{MeV}\else $\operatorname{MeV}$\fi}
\protected\def\GeV{\ifmmode \,\operatorname{GeV}\else $\operatorname{GeV}$\fi}
\protected\def\TeV{\ifmmode \,\operatorname{TeV}\else $\operatorname{TeV}$\fi}

\protected\def\Lsun{\ifmmode \,\operatorname{L_\odot}\else $\operatorname{L_\odot}$\fi}
\protected\def\lsun{\ifmmode \,\operatorname{L_\odot}\else $\operatorname{L_\odot}$\fi}
\protected\def\Watt{\ifmmode \,\operatorname{W}\else $\operatorname{W}$\fi}
\protected\def\nW{\ifmmode \,\operatorname{nW}\else $\operatorname{nW}$\fi}

\protected\def\kJy{\ifmmode \,\operatorname{kJy}\else $\operatorname{kJy}$\fi}
\protected\def\Jy{\ifmmode \,\operatorname{Jy}\else $\operatorname{Jy}$\fi}
\protected\def\mJy{\ifmmode \,\operatorname{mJy}\else $\operatorname{mJy}$\fi}
\protected\def\microJy{\ifmmode \,\operatorname{\mu Jy}\else $\operatorname{\mu Jy}$\fi}
\protected\def\nJy{\ifmmode \,\operatorname{nJy}\else $\operatorname{nJy}$\fi}

\protected\def\Mag{\ifmmode \,\operatorname{mag}\else $\operatorname{mag}$\fi}

\protected\def\deg{\ifmmode ^{\circ}\else $^{\circ}$\fi}
\protected\def\arcsec{\ifmmode ^{\prime\prime}\else $^{\prime\prime}$\fi}

\protected\def\arcsecT{\ifmmode \,\operatorname{arcsec}\else $\operatorname{arcsec}$\fi}
\protected\def\arcmin{\ifmmode ^{\prime}\else $^{\prime}$\fi}

\protected\def\arcminT{\ifmmode \,\operatorname{arcmin}\else $\operatorname{arcmin}$\fi}
\protected\def\sr{\ifmmode \,\operatorname{sr}\else $\operatorname{sr}$\fi}

\protected\def\d{\ifmmode \operatorname{d}\else
    $\operatorname{d}$\fi}
\protected\def\e{\ifmmode \operatorname{e}\else
    $\operatorname{e}$\fi}

\def\apjl{{ApJ}}

\def\apjs{{ApJS}}

\def\apjref#1;#2;#3;#4 {\par\pp#1, {#2}, #3, #4 \par}

\shorttitle{Luminosity Functionals}
\shortauthors{Lake et al.}

\begin{document}

\title{The 2.4 $\mu$m Galaxy Luminosity Function as Measured Using \WISE. I. Measurement Techniques}
\author{S.~E.~Lake\altaffilmark{1}, E.~L.~Wright\altaffilmark{1}, C.-W.~Tsai\altaffilmark{1,2}, A.~Lam\altaffilmark{1}}

\altaffiltext{1}{Physics and Astronomy Department, University of California, Los Angeles, CA 90095-1547}
\altaffiltext{2}{Jet Propulsion Laboratory, California Institute of
Technology, 4800 Oak Grove Dr., Pasadena, CA 91109}

\email{lake@physics.ucla.edu}

\begin{abstract}
The astronomy community has at its disposal a large back catalog of public spectroscopic galaxy redshift surveys that can be used for the measurement of luminosity functions. Utilizing the back catalog with new photometric surveys to maximum efficiency requires modeling the color selection bias imposed on selection of target galaxies by flux limits at multiple wavelengths. The likelihood derived herein can address, in principle, all possible color selection biases through the use of a generalization of the luminosity function, $\Phi(L)$, over the space of all spectra: the spectro-luminosity functional, $\Psi[L_\nu]$. It is, therefore, the first estimator capable of simultaneously analyzing multiple redshift surveys in a consistent way. We also propose a new way of parametrizing the evolution of the classic Schechter function parameters, $L_\star$ and $\phi_\star$, that improves both the physical realism and statistical performance of the model. The techniques derived in this paper are used in a companion paper \cite{Lake:2017c} to measure the luminosity function of galaxies at the rest frame wavelength of $2.4\micron$ using the \textit{Widefield Infrared Survey Explorer} (\WISE).
\end{abstract}



\section{Introduction}
The luminosity function (LF), which describes the demography of objects with respect to their luminosity, is one of the most basic statistical properties that can be measured for objects observed in astronomy. Because it is so basic, and because luminosities cannot be measured directly, the number of ways to estimate the LF for galaxies are numerous \citep[see reviews:][]{Johnston:2011, Willmer:1997, Binggeli:1988}; so numerous, in fact, that it is traditional for papers measuring a luminosity function for galaxies to use multiple estimators, at least one parametric and one non-parametric. The most commonly used non-parametric estimators (refinements of the $1/V_{\mathrm{max}}$ estimator of \cite{Schmidt:1968} made by \cite{Avni:1980} and step-wise maximum likelihood [SWML] from \cite{SWML}) can be accurately described as a binning of the data with completeness corrections. The most popular parametric estimators are variations on the one from \cite{STY}, commonly called STY.

This work seeks to address three weaknesses in how parametric model fitting is currently done for the purpose of applying it to real data in the companion paper \citep[LW17III]{Lake:2017c}. First, the most common method for addressing incompleteness is not likelihood based and is, therefore, likely to converge more slowly and exhibit higher bias than a more strictly likelihood based one. Second, the existing techniques in the literature can only cope with data sets that are volume limited, flux limited in a single bandpass, or that otherwise have a sharp cutoff in luminosity-redshift space and a constant observation selection probability. Third, the majority also do not address a bias related to spectral energy distribution variety, explored in \cite{Ilbert:2004}. 

This paper includes the derivation of an estimator that addresses all of these concerns and can, in principle, handle an arbitrary number of flux selection limits. Each flux limit requires an additional dimension in the numerical integration of some probability density function (PDF); for this work that PDF is a multi-dimensional Gaussian. The well known ``curse of dimensionality" places limits on the performance of such deterministic numerical integrations, so the estimator is presently useful for analyzing surveys with flux limits in at most two bands. The basis of this new estimator is rooted in a generalization of the luminosity function that uses functional analysis (calculus of variations) to estimate the density of galaxies per unit volume, per unit luminosity, per unit spectral energy distribution (SED) function space volume. This generalization is most naturally called the spectro-luminosity functional.

The derivation is included to also show the way that many of the most popular estimators are related to the one defined in this work. Of the parametric estimators, STY and the estimator from \cite{Marshall:1983} are approximations of the one from this work. The following binned LF estimators are also related to the estimator defined in this work: $1/V_{\mathrm{max}}$ of \cite{Schmidt:1968}, the binned Poisson estimator of \cite{Page:2000}, and corrected Poisson estimator of \cite{Miyaji:2001}.

It has been observed in multiple measurements of the LF that the density ($\phi_\star$) and luminosity scale ($L_\star$) parameters are evolving with time \citep{Lin:1999,Blanton:2003LF,Babbedge:2006,Dai:2009,Loveday:2012,Cool:2012}. The most frequently used parametrizations for the evolution of $\phi_\star$ and $L_\star$ are the ones from \cite{Lin:1999}: $L_\star \propto 10^{0.4 Q z}$ and $\phi_\star \propto 10^{0.4 P z}$, with $Q$ and $P$ constants. These parametrizations work well at the empirical level, as long as the evolution is not extrapolated to very high redshifts where, depending on the measured values of the evolution constants $P$ and $Q$, either or both $\phi_\star$ and $L_\star$ can become un-physically large at high redshifts. To remedy this, we propose a similar modified parametrization based on lookback time, instead of redshift, that also satisfies the boundary condition that at some point in the past $L_\star$ was zero.

There are a number of physical parameters that are derivable from the LF, including: galaxy number density (brighter than some cutoff), the specific rate of change in number density, luminosity density, the specific rate of change in luminosity density, and the normalization of the predicted histogram of extragalactic sources binned by observed flux. This work contains formulae that relate these parameters to the Schechter parameters, with specific application to the form they take when applied to the evolution scheme from \cite{Lin:1999} and the one proposed here. There is also a brief discussion of which parameters, from experience, give less correlated errors when used in a Markov Chain Monte Carlo (MCMC) characterization of the Bayesian posterior of the maximum likelihood estimator from this work.

The structure of this paper is as follows: Section~\ref{sec:measure} contains derivations of multiple ways of measuring LFs based on the likelihood, using both binned and unbinned techniques, Section~\ref{sec:Parameterization} covers different aspects of describing the evolution of Schechter LF parameters, offering improvements motivated on both physical grounds and statistical performance, and Section~\ref{sec:discussion} contains a brief discussion on where there is room for improvements going forward. The performance of the estimators on simulated data sets is not examined, because they are 100\% likelihood based, and any simulations that can be done quickly would require the same approximations made in deriving the likelihood. Instead, the tools are applied to real world data in LW17III to measure the luminosity function of galaxies at rest frame $2.4\micron$ using \WISE\ in combination with a number of public spectroscopic redshift surveys. 

\section{Luminosity Function Measurement} \label{sec:measure}
The basic definition a luminosity function, $\Phi(L)$, is the number density of sources per unit luminosity per unit volume. This definition often leads to the treatment of $\Phi$ as a PDF that is normalized to the density of sources instead of unity. For galaxies this treatment conflicts with the fact that the integral of the reported luminosity functions is frequently divergent \citep[see, for example, the results in][]{Dai:2009}. The contents of the section break down, by subsection, as follows: Subsection~\ref{sec:msr:density} resolves this conflict by more precisely defining $\Phi$ as a non-normalized density, permitting a rigorous derivation of a corresponding maximum likelihood estimator. Subsection \ref{sec:msr:slf} covers the technique used here to deal with the biases that cosmological redshifting adds to the problem of measuring galaxy luminosity functions over large redshift ranges. The technique used is to expand the definition of luminosity functions to cover their entire spectrum: the spectro-luminosity functional, $\Psi[L_\nu]$. Section~\ref{sec:msr:bin} derives the corresponding maximum likelihood estimator for a binned LF estimator, showing how different estimators in the literature correspond to different levels of approximation. Finally, Subsection~\ref{sec:msr:err} gives formulae for estimating the uncertainties in the binned estimators described in the previous subsection.

\subsection{Luminosity Function Definition}\label{sec:msr:density}
In this work we choose to define $\Phi$ rigorously, bordering on pedantically, in order to show where all of the approximations that leave room for future refinements are, and to show how where many existing estimators are approximations of the one derived here. The definition of $\Phi$ used here is as follows: given a volume, \d$V_i$, around the point in space at position $\vec{x}_i$, and a luminosity interval, \d$L_i$, around the luminosity $L_i$, the average number of sources that exist in that bin around that point in luminosity-location space is:
\begin{align}
	\langle m_i \rangle & = \Phi(L_i,\, \vec{x}_i) \d L_i \d V_i, \label{eqn:thry:LFbase}
\end{align}
where the average is taken over an ensemble of universes with different initial conditions, $m_i$ is the actual number of galaxies around $(L_i,\, \vec{x}_i)$ in a particular universe selected from the ensemble, and $i$ is an index that covers all space-luminosity boxes in a given partition scheme of a particular universe. The homogeneity and isotropy of the universe guarantees that $\Phi$ does not actually depend on $\vec{x}$, but it can evolve with time, and, therefore, does depend on redshift, $z$. What differentiates this definition from others is that it is explicitly an average over a hypothetical ensemble of universes, making it conceptually independent of environment. A similar definition can be arrived at for an environmentally dependent LF, what we would call a conditional luminosity function, the two most common examples of which are the field galaxy LF (low density environment) and the cluster galaxy LF (high density). However, doing so introduces complications to the measurement process related to correctly identifying the environment each galaxy is in, and how to correctly match those classifications across cosmological time, that are beyond the scope of this work.

We can assume that $m_i$ is Poisson distributed, an approximation that is accurate in the limit where the volume of the spatial box is $\langle m_i \rangle \ll 1$, as is done implicitly in \cite{Marshall:1983}. The random vagaries of whether the or not the galaxies in box $i$ are selected in the observation process can be modeled as a binomial process with probability denoted $S(L_i, \vec{x}_i)$, called the selection function, and a number of trials equal to $m_i$. When the binomial observation process is combined with the Poisson distribution of $m_i$, and the number of unobserved galaxies is summed over, the resulting distribution of galaxies observed, $n_i$, has probability:
\begin{align}
	P(n_i) & = \frac{\left[ S_i \Phi_i \d L_i \d V_i\right]^{n_i}}{n_i!} \mathrm{e}^{- S_i \Phi_i \d L_i \d V_i}. \label{eqn:thry:pois}
\end{align}

It is in the next step that the biggest approximation is made. The approximation is to treat all of the $P(n_i)$ as independent, as implied by the cosmological principle (homogeneity and isotropy of the universe). Where this fails is that galaxies have finite size, and overlapping galaxies tend to quickly merge, so no more than one galaxy of any luminosity can occupy a given region of space for long, and the size of that region depends on the luminosity of the galaxy. Outside of that effectively excluded volume, however, galaxies tend to be positively correlated (that is, the presence of a galaxy at one point increases the probability of finding another galaxy nearby), as measured in works on galaxy clustering \citep[for example:][]{Zehavi:2005}. A full treatment of the impact these factors would have on the true likelihood of galaxy catalogs is beyond the scope of this work. The cosmological principle approximation applied to Equation \ref{eqn:thry:pois} implies that:
\begin{align}
	\ln(P_{\mathrm{catalog}}) & = \sum_i \left[ n_i \ln\left( S_i \Phi_i \d L_i \d V_i\right) - \ln(n_i!) \right. \nonumber \\
	&\hphantom{= \sum_i [} \left. - S_i \Phi_i \d L_i \d V_i\right], 
\end{align}
where $P_{\mathrm{catalog}}$ is the probability of observing the catalog of galaxies in the bins being summed over.

Taking the limit that the space-luminosity boxes are small enough that all $n_i$ are either 0 or 1 reduces the first term of the sum to be only over those boxes where a galaxy was observed, the second term vanishes, and the third term becomes an integral. Also trading the probability, $P_{\mathrm{catalog}}$, for a likelihood, $\mathcal{L}_{\mathrm{catalog}}$, gives the penultimate form of the maximum likelihood estimator:
\begin{align}
	\ln(\mathcal{L}_{\mathrm{catalog}}) & = \sum_i \ln\left( S(L_i, \vec{x}_i)\, \Phi(L_i,\, z_i) \right) \nonumber \\
	&\hphantom{=} - \int S(L, \vec{x})\, \Phi(L,\, z) \d L \d V. \label{eqn:thry:estComb}
\end{align}
In the sharply cutoff step function completeness limit, that is $S(L, \vec{x})$ takes on one of two constant values, $0$ and $s$, Equation \ref{eqn:thry:estComb} becomes identical to the estimator from \cite{Marshall:1983}.

The next refinement to the estimator is to isolate the normalization of $\Phi$. That is, if $\Phi = \phi_\star \Phi_0(L, z)$, then the likelihood splits into a maximum likelihood estimate for $\phi_\star$ and a likelihood for $\Phi_0$:
\begin{align}
	\phi_\star & = \frac{N}{\int S(L, \vec{x})\, \Phi_0(L,\, z) \d L \d V},\ \mathrm{and} \\
	\ln(\mathcal{L}_{\mathrm{catalog}}) & = \sum_i \ln\left( \frac{S(L_i, \vec{x}_i)\, \Phi_0(L_i,\, z_i)}{\int S(L, \vec{x})\, \Phi_0(L,\, z) \d L \d V} \right), \label{eqn:thry:estSTY}
\end{align}
where $N$ is the total number of galaxies in the catalog, and terms that in the sum that do not depend on the parameters in $\Phi$ are dropped as irrelevant. If the selection function, $S(L_i, \vec{x}_i)$, depends only on the observed flux and the differing effect of cosmological redshifting on galaxies with different SEDs is not significant, then Equation~\ref{eqn:thry:estSTY} becomes identical to the estimator in \cite{STY}, called the STY estimator for the authors of the paper. 

The original definition of the estimator given in \cite{STY} handled incompleteness in a similar way to what we do here, with a more simplified model for it. Some time between the review by \cite{Binggeli:1988} and the work of \cite{Efstathiou:1988} the handling of incompleteness changed to weighting the terms of the log-likelihood sum by the inverse of the selection function, as described in Equation~27 of the review of \cite{Johnston:2011}. Weighting terms in a likelihood estimator with real valued scalars is identical to placing those same scalars into the exponents of the product of the likelihoods. This can make a certain amount of intuitive sense, if the weighting is viewed as correcting the number of observations at each point to be what would have been observed were the data set complete. 

\begin{figure}[htb]
	\begin{center}
	\includegraphics[width=0.45\textwidth]{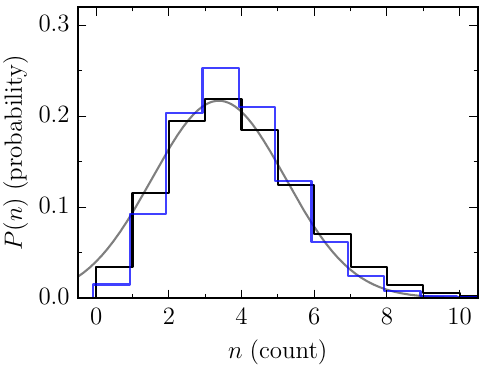}
	\end{center}
	
	\caption{ Comparison of the distribution of a true Poisson variable (black histogram) to a ``completeness corrected" version that has the same mean (blue histogram, offset left) and a Gaussian distribution with matching mean and variance (grey line). The black line has mean $\lambda = 4.5$ and completeness $s=0.75$, for a combined mean of $3.375$. The blue line is the result of taking the formula for a Poisson distributed variable, $P(n) = \lambda^n \e^{-\lambda} / n!$, completeness correcting it with $n \rightarrow n / s$, and normalizing the distribution to sum to unity. The blue distribution, which is what current completeness correction techniques implicitly use, is narrower than the black, underestimating the variance of the distribution. The Gaussian distribution with mean and variance $3.375$ is provided to show that it more closely matches the black histogram than the blue one does. }
	\label{fig:thry:Poisson}
\end{figure}

The problem with viewing the use of weights in the sum of log-likelihoods as correcting the data is that it does not model the statistical behavior of the completeness corrected data, making estimates based on the weighted sum not likelihoods. While a general demonstration for any binned data set of independent and identically distributed data is possible, the steps are fundamentally the same as if we consider a single Poisson distributed quantity, $n$, with mean $s \lambda$. In this case, the completeness corrected data, $n/s$, will have mean $\lambda$ and variance $\lambda / s$. Just dividing the log-likelihood by $s$, as described, treats $n/s$ as though it were Poisson distributed with mean and variance $\lambda$. Figure~\ref{fig:thry:Poisson} shows a comparison between the distribution of a Poisson distributed quantity, $n$, with $\lambda = 4.5$ and $s=0.75$ in comparison with how the weighted log-likelihood behaves with $n$ and a Gaussian with matching mean and variance.
A detailed analysis of how getting the variance, and all higher order moments, wrong affects the performance of the estimator is beyond the scope of this work, however it is not a stretch to say that using a distribution that gets the mean right but the variance wrong is an approximation that is not even Gaussian level, which gets the mean and variance right, and therefore likely to exhibit a greater bias than a correct likelihood treatment is. It is worth noting that we are not claiming that working with completeness corrected data is not possible, just that doing so should be limited to cases where the number of objects pre-correction in each bin is large enough that the bin's distribution can be approximated with a Gaussian (the limit where $\chi^2$ statistics are a good approximation), permitting correct treatment of both the bin's mean and variance.

The integral in Equations \ref{eqn:thry:estComb}--\ref{eqn:thry:estSTY} is often only evaluable numerically. As long as the numerical integration only has to be performed once per calculation of $\mathcal{L}_{\mathrm{catalog}}$, and not for every term in the sum, then the computation is usually not too computationally expensive. With both of Equations~\ref{eqn:thry:estComb} and \ref{eqn:thry:estSTY}, though, the selection function for a flux-limited sample will be different for each source because each source has a unique SED, thus requiring the numerical integral to be evaluated for each source in the catalog. It is the ability to avoid this burdensome calculation that makes the more advanced estimator derived in Section~\ref{sec:msr:slf} useful.

\subsection{The Spectro-Luminosity Functional, $\Psi$}\label{sec:msr:slf}
\begin{figure*}[htb]
	\begin{center}
	\includegraphics[width=\textwidth]{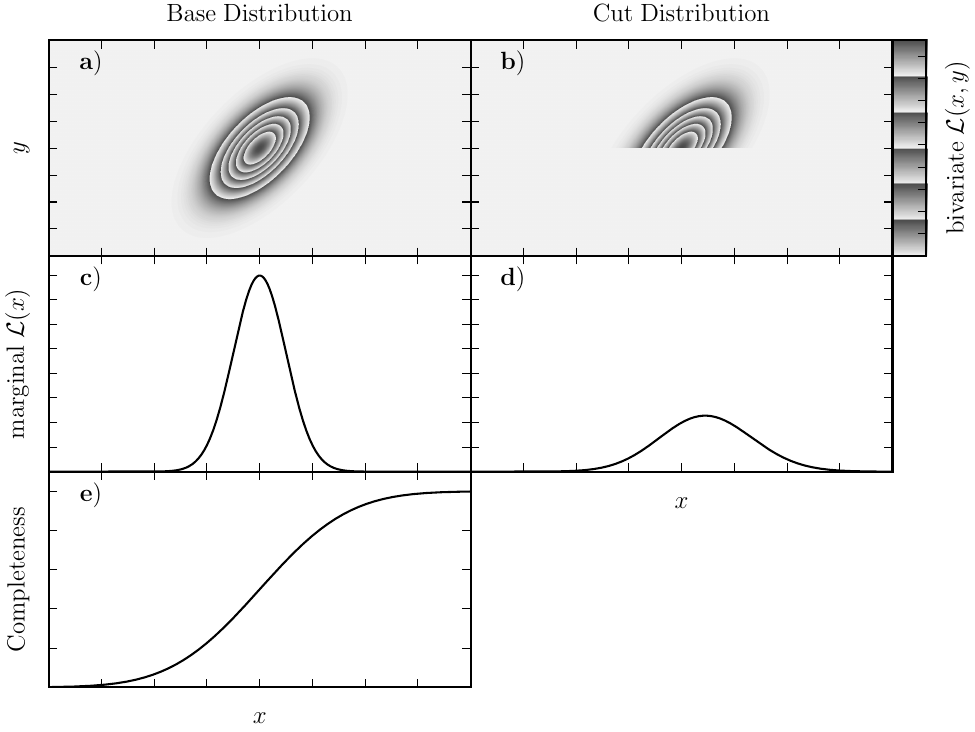}
	\end{center}
	\caption{Illustration how a sharp cut on one variable imposes a blurred cut on a correlated variable. The relationships are simplified, but the concept is the same for the relationship between flux and luminosity with SED variability thrown in. Panel~\textbf{a} contains the base, uncut, bivariate distribution in $x$ and $y$. Panel~\textbf{b} shows the same distribution as \textbf{a} with a cut imposed based on $y$ with the same color scale. Panel~\textbf{c} shows the projected (marginal) distribution in $x$ obtained from integrating Panel~\textbf{a} over all $y$. Panel~\textbf{d} shows the projected (marginal) distribution in $x$ obtained from integrating Panel~\textbf{b} over all $y$ on the same scale as Panel~\textbf{c}. Panel~\textbf{e} shows the completeness of the distribution in \textbf{d} with respect to the uncut distribution in \textbf{c} (the ratio of \textbf{d} over \textbf{c}).
	}
	\label{fig:thry:SharpCut}
\end{figure*}

The phenomenon of cosmological redshifting combined with the variety of galaxy SEDs complicates the process of translating flux selection limits into luminosity selection limits, even for a single band selected survey. In the absence of redshifting, a sharp cut along a straight line in a flux-distance graph translates into a curved line in a luminosity-distance graph, but the sharpness of the cut boundary is unaffected. If all galaxies had the same SED, then the statement in the previous sentence would be unmodified because it would just mean that the boundary curve would have a different shape. For any real collection of galaxies, though, each galaxy has a different SED, making the relationship between flux and luminosity for galaxies overall effectively statistical. Figure~\ref{fig:thry:SharpCut} illustrates a simplified situation in which the fundamental mechanisms are the same: a cut in one variable imposes a smoothly varying completeness in a correlated variable. It is, therefore, necessary to model the distribution of the variety of galaxy SEDs to accurately reproduce the selection function needed by the maximum likelihood estimator. 

The straightforward approach to modeling the distribution of SEDs is to generalize the concept of the luminosity function beyond a density of galaxies per unit luminosity, as measured by a single detector, to the density of galaxies per unit volume in the space of possible SEDs, that we are calling $\Psi$. One advantage of this approach is that it provides a universal intermediate form between models that predict the distribution of galaxy SEDs and all of the different histograms astronomers construct from galaxy luminosities and fluxes (especially: luminosity functions and color-magnitude diagrams). Most importantly, because those predicted histograms are all constructed from the same intermediate, data from disparate surveys with different selection criteria can feed information back into the theory's parameters in a consistent way. 

This section takes the following approach: define $\Psi$ and its relation to the standard LF, construct the part of $\Psi$ that extends beyond $\Phi$ using a Gaussian approximation, and define an approximation for the Gaussian's parameters using SED templates to reduce the parameters from full functions to matrices. 

The full details of how probability densities of functions are handled can be found in some texts on functional analysis, or any text on quantum mechanics or quantum field theory that covers the path integral approach \citep[for example:][]{PeskinSchroeder}. In a broad sense, the calculus of functions is the same as ordinary vector calculus because a function is as an element of a vector space; the only distinguishing feature is that a limit as the number of dimensions approaches infinity is implicitly taken. 

\begin{figure}[htb]
	\begin{center}
	\includegraphics[width=0.48\textwidth]{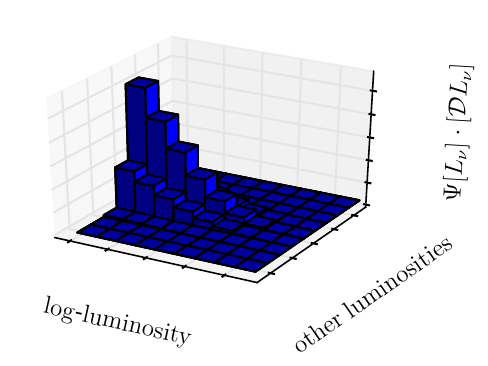}
	\end{center}
	\caption{Cartoon illustrations of a function space density with the functions approximated by a sequence of $N$ step functions. The $x$ and $y$ axes represent some measure of a galaxy's luminosity and all the other degrees of freedom in the SED, respectively. The value of $\Phi(L) \operatorname{d}L$ can be found by summing over the uncountably infinite number of axes hidden in the ``other luminosities'' axis. 
	}
	\label{fig:thry:funcspace}
\end{figure}

The definition of $\Psi$ begins similarly to the definition of $\Phi$ given in the previous section. The mean number of galaxies within a function space volume, $[\mathcal{D} L_\nu]_i$, of the SED $L_\nu(\nu)$, and within a real space volume $\d V_i$ of $\vec{x}_i$ is given by:
\begin{align}
	\langle n \rangle & = \Psi[L_\nu](z_i)\cdot \d V_i \cdot [\mathcal{D} L_\nu]_i, \label{eqn:thry:basePsi}
\end{align}
where the square braces denote that $\Psi$ is a functional of $L_\nu$ and the parentheses denote that it is an ordinary function of $z_i$. 
A non-rigorous illustration of a density over function space, the concept embodied in Equation~\ref{eqn:thry:basePsi}, can be found in Figure~\ref{fig:thry:funcspace}. The lack of specificity in defining what is meant by a ``function space volume" is actually reflective of the current state of the field of functional integration \citep[see, for example:][and references therein]{Cartier:2000, Nakahara, Zinn-Justin:2009, Albeverio:2011}.

The relationship between $\Psi[L_\nu](z)$ and $\Phi(L, z)$ is a marginalization (that is, summing over irrelevant degrees of freedom):
\begin{align}
	\Phi(L, z) & = \int [\mathcal{D} L_\nu]\, \delta\left(L - \int L_\nu(\nu) w(\nu, z)\d \nu \right)\, \Psi[L_\nu](z),
\end{align}
where $w(\nu, z)$ is a weighting function that, through its units and form, defines what the symbol $L$ means, and $\delta(x)$ is the Dirac delta function. For example, if the weighting function depends on redshift in the following way then $L$ is an observer frame flux ready for conversion to a magnitude:
\begin{align}
	w(\nu, z) & = \frac{(1+z)\, \nu^{-1}\, R(\nu / [1 + z])}{4\pi D_L(z)^2 \int R(\nu)\,F_{\nu,\mathrm{std}}(\nu)\, \nu^{-1} \d \nu},
\end{align}
where $F_{\nu,\mathrm{std}}(\nu)$ is a flux SED of some standard source, $D_L(z)$ is the luminosity distance, and $R(\nu)$ is the relative detector response to a photon with observer frame frequency $\nu$. As the example equation suggests, the weight functions are meant as a generalization of detector responses, regardless of whether the response is: narrow, as it is for a pixel in a spectrograph; broad, as for an imaging filter; or fully bolometric. Performing similar marginalizations with multiple different weighting functions can produce any color-magnitude diagram, up to a change of variables and application of a selection function, in observer frame or rest frame quantities.

\subsection{$\Psi$ in Gaussian Approximation}\label{sec:msr:slfG}
The use of the calculus of functions makes $\Psi$ extremely general and powerful conceptually, however concrete predictions require a projection down from a density over infinite-dimensional function space to one over a finite number of dimensions (a process called marginalization in statistics). The algorithms for doing this numerically for the general case are computationally intensive (see, for example, the field of study known as lattice quantum chromodynamics), so it is important to find a form for $\Psi$ that can be marginalized analytically. To begin with, in order to make the form of $\Psi$ consistent with what is already known about $\Phi$, it is useful to use the ratio of $\Psi$ and $\Phi$ to define the conditional likelihood functional of the SED ($\mathcal{L}_{SED}$):
\begin{widetext}
\begin{align}
	\Psi[L_\nu](z) & = \Lsed\left[\hat{\Pi}_\perp L_\nu \right](.|L, z) \times \Phi(L, z), \label{eqn:thry:LsedDef}
\end{align}
\end{widetext}
where $\Lsed\left[\hat{\Pi}_\perp L_\nu \right](.|L, z)$ symbolizes that \Lsed\ is a functional of the random function $\hat{\Pi}_\perp L_\nu$ and a function of the non-random variables $L$ and $z$, with the dot there as a placeholder because there are no random scalars on which \Lsed\ depends. $\hat{\Pi}_\perp$ is an operator that adjusts the SED, $L_\nu$, to satisfy one or more constraints equations of the form:
\begin{align}
	\int \left[\hat{\Pi}_\perp L_\nu \right](\nu) \times w(\nu, z) \d \nu &= L.
\label{eqn:thry:constraint}
\end{align}
When there is a single constraint, $\hat{\Pi}_\perp$ can be constructed as a normalization of the SED:
\begin{align}
	\hat{\Pi}_\perp L_\nu & \equiv L \frac{L_\nu}{\int L_\nu(\nu)\, w(\nu) \d \nu }.
\end{align}
This is not the only way to define $\hat{\Pi}_\perp$. Factors that affect the form of $\hat{\Pi}_\perp$ include: the number of constraints, whether the weight functions overlap, and whether it is important to maintain the non-negativity of the allowed SEDs. If the form of \Lsed\ assigns sufficiently low likelihood to un-physically negative SEDs, and all of the weight functions are linearly independent, for example, it is possible to treat $L_\nu$ as a vector and construct $\hat{\Pi}_\perp$ as a linear projection operator by performing Gramm-Schmidt orthogonalization on the weight functions.

A more accurate form of \Lsed\ is derivable using the star formation history of the universe, stellar population synthesis models, dust models, active galactic nuclei (AGN) accretion history models, and some description of the density fluctuations in the universe. The last is important because galaxies with red SEDs, indicating ``quiescent" star formation, are typically found in higher density environments than galaxies with the optically blue SEDs that indicate active star formation, so density fluctuations will affect the spread of possible SEDs and balance of red and blue types. This work avoids those complications by just approximating \Lsed\ as a single Gaussian functional, with the understanding that the approximation can be improved by either combining multiple Gaussians in \Lsed\ or summing over different luminosity functionals for different types of galaxy (for example, $\Psi_{\mathrm{overall}} = \Psi_{\mathrm{AGN}} + \Psi_{\mathrm{red}} + \Psi_{\mathrm{blue}}$). Explicitly, let $w_0(\nu)$ be the weight function that defines the rest frame luminosity that is the argument of $\Phi$, and $\ell_\nu \equiv L_\nu / \int L_\nu w_0 \d \nu$, then:
\begin{widetext}
\begin{align}
	\vphantom{-}\Lsed[\ell_\nu] & = \frac{\mathcal{N}}{\sqrt{\mathrm{Det}[\Sigma]}}  \nonumber \\
		&\hphantom{=}\cdot \exp\left(-\frac{1}{2} \int \d \nu \d \nu' \left[\ell_\nu(\nu) - \mu_\nu(\nu)\right]
		 \Sigma^{[-1]}(\nu, \nu') \left[\ell_\nu(\nu') - \mu_\nu(\nu')\right]\right). \label{eqn:thry:LsedGaus}
\end{align}
\end{widetext}
$\mu_\nu(\nu)$ is the mean of the normalized SEDs, $\Sigma(\nu, \nu')$ is the covariance of normalized SEDs, and $\mathcal{N}$ is a normalization factor with a form that depends on the details of how the function space is parametrized. If the functions are approximated with $M$ boxcars, for example, $\mathcal{N}= (2\pi)^{-M/2}$. $\Sigma^{[-1]}(\nu, \nu')$ is the functional inverse of $\Sigma(\nu, \nu')$, that is it satisfies: $\int \Sigma(x, y) \Sigma^{[-1]}(y, z) \d y = \delta(x - z)$. Perhaps the most familiar example of a functional inverse in physics is that of the negative Laplacian operator ($-\nabla^2$), called its Green's function, given by:  $G(\vec{r}, \vec{r}') = 1/(4\pi |\vec{r} - \vec{r}'|)$. Another example specifically from astronomy is implicitly used in the process of image deconvolution; it is formally the same as treating the image's point spread function as the kernel of a linear operator and finding, approximately, that operator's inverse.

It may seem like using a single Gaussian for \Lsed\ is too great an approximation because it cannot produce the most prominent structures seen in color-magnitude diagrams of galaxies: the ``red sequence" and ``blue cloud". Doing so is equivalent to a standard practice of performing a single LF fit to all of the data to produce an overall LF \citep[for example:][]{Efstathiou:1988,Loveday:2000,Cole:2001,Kochanek:2001,Blanton:2003LF,Bell:2003,Jones:2006,Babbedge:2006,Cirasuolo:2007,Dai:2009,Smith:2009,Montero:2009,Loveday:2012,Cool:2012}. A further level of refinement is usually, but not always, included where the galaxies are broken down by type (most frequently: early/red ``quiescent" galaxies, late/blue ``star forming" galaxies, and AGN). This would be equivalent to treating $\Psi$ as the sum of three spectro-luminosity functionals that each have one Gaussian in \Lsed.

The mean of the normalized SEDs, $\mu_\nu$, is required to compute the spectral comoving luminosity density (CLD) of the cosmos, $\rho_{L_\nu}$; it is
\begin{align}
	\rho_{L_\nu}(\nu) & = \int [\mathcal{D}L_\nu]\,  L_\nu\, \Psi[L_\nu](z)\nonumber \\
	& = \mu_\nu(\nu) \int L\, \Phi(L, z) \d L.
\end{align}
If $\mu_\nu$ depends on luminosity, as is allowed, then it will be inside the integral.

Returning to the derivation of an LF estimator from Equation~\ref{eqn:thry:LsedGaus}, the Gaussian form of \Lsed\ and the linearity of detectors makes the transition from function space to a finite number of bandpass fluxes and luminosities an exercise in linear algebra that introduces no new approximations. If there are $K$ luminosities/fluxes of interest, defined by weighting functions $w_k(\nu, z)$ with $k=1\ldots K$, then:
\begin{align}
	\vphantom{-}\Lsed(\{\ell_k\}) &= \frac{1}{\sqrt{(2\pi)^K \operatorname{det}(\Sigma)}} \nonumber \\
	&\hphantom{=} \cdot \exp\left(-\frac{1}{2} \sum_{i, j=1}^K [\ell_i - \mu_i] [\Sigma^{-1}]_{ij}[\ell_j - \mu_j] \right), \label{eqn:thry:mLL}
\end{align}
where $\ell_k \equiv \int \ell_\nu(\nu) w_k(\nu, z) \d \nu$, $\mu_k \equiv \int \mu_\nu(\nu) w_k(\nu, z) \d \nu$, \break 
$\Sigma_{ij} \equiv \int \Sigma(\nu, \nu') w_i(\nu, z) w_j(\nu', z) \d \nu \d \nu'$, and projecting down from function space to this $K$-dimensional space has transformed the functional determinant and inverse into ordinary matrix ones. Equation~\ref{eqn:thry:mLL} works as long as all of the weighting functions, $w_k(\nu, z)$, are linearly independent functions.  

In principle, any number of fluxes can be accommodated in \Lsed, but every flux that has a sensitivity limit imposed on it will add a dimension to a numerical integration that must be performed every time the full likelihood is calculated. Because of this, we specialize to the case where $K=2$, where fast analytical calculations of the integral are still possible.

Equation~\ref{eqn:thry:LsedGaus} is an approximation because physical SEDs are all non-negative, and the Gaussian form in the equation does not exclude un-physically negative ones. That constraint could have been satisfied by choosing a log-Normal form for the distribution instead of Gaussian, but then the projection into a subspace using a set of weighting functions becomes only possible in terms of spectral quantities, that is $w_k \propto \delta(\nu - \nu_k)$. As long as the value of the mean exceeds the standard deviation by enough, for example $\mu_\nu(\nu) > 2 \sqrt{\Sigma(\nu, \nu)}$, for all relevant frequencies the approximation should be accurate enough. Finding a more correct form for \Lsed\ that satisfies this constraint without approximation is left for future work.

\subsection{Approximating the Mean and Variance of SEDs Using Templates}\label{sec:msr:tmplts}
In principle, $\mu_\nu$ and $\Sigma$ can be measured directly using the population mean and covariance of directly observed, normalized, spectra:
\begin{align}
	\mu_\nu(\nu) & \approx \frac{1}{N}\sum_{i=1}^N \ell_{\nu,\,i}(\nu) \equiv \overline{\ell_\nu}(\nu),\ \mathrm{and} \nonumber \\
	\Sigma(\nu, \nu') & \approx \frac{1}{N-1} \sum_{i=1}^N (\ell_{\nu,\,i}(\nu) - \overline{\ell_\nu}(\nu))(\ell_{\nu,\,i}(\nu') - \overline{\ell_\nu}(\nu')). \label{eqn:thry:meancov1}
\end{align}
If sufficient high resolution spectroscopic data were available to actually measure $\mu_\nu$ and $\Sigma$ as functions, then it would not be necessary to construct $\Psi$ to measure $\Phi$, as $\Phi$ could be measured directly. It is, therefore, necessary to further approximate the galaxy SEDs as a sum over a set of templates, for example the four templates of \cite{Assef:2010}. If there are $N_{\mathrm{T}}$ templates in the set with SEDs $\tilde{\ell}_{\nu,\,a}$, then under the template approximation for each galaxy $\ell_{\nu, \,i}(\nu) \approx \sum_{a=1}^{N_{\mathrm{T}}} f_{ai}\ \tilde{\ell}_{\nu,\, a}(\nu)$. Applying the template approximation to Equations~\ref{eqn:thry:meancov1} yields:
\begin{align}
	\mu_\nu(\nu) & \approx \sum_{a=1}^{N_{\mathrm{T}}} \overline{f_a}\, \tilde{\ell}_{\nu,\, a}(\nu),\ \mathrm{and} \nonumber\\
	\Sigma(\nu, \nu') & \approx \sum_{a, b=1}^{N_{\mathrm{T}}} \operatorname{cov}(f_a, f_b)\, \tilde{\ell}_{\nu,\, a}(\nu)\, \tilde{\ell}_{\nu,\, b}(\nu'),\label{eqn:thry:meancov2}
\end{align}
where $\overline{f_a}$ and $\operatorname{cov}(f_a, f_b)$ are a vector and matrix of parameters, respectively. It is worth reiterating that $\tilde{\ell}_{\nu,\, a}$ are the normalized template spectra, and $f_{ai}$ is the fraction of luminosity contributed by template $a$ to the $w_0$ band pass for the galaxy identified by index $i$.

It is possible to include the parameters in Equations~\ref{eqn:thry:meancov2}, $\overline{f_a}$ and $\operatorname{cov}(f_a, f_b)$, with the others when fitting the LF to data. Doing so greatly increases the number of free parameters in the fit, however, with $N_T-1$ degrees of freedom in $\overline{f_a}$ and $(N_T-1)(N_T-2) / 2$ degrees of freedom in $\operatorname{cov}(f_a, f_b)$, consistent with the normalization condition on the luminosity fractions. It is, therefore, worth it to fix these parameters separately from the measurement of the luminosity function, if possible.

The final form of the LF likelihood used in LW17III can now be written down by combining Equations~\ref{eqn:thry:estComb}, \ref{eqn:thry:LsedDef}, and \ref{eqn:thry:mLL}. Note that we introduce a new selection function here that plays the role of the sharp cut between panels \textbf{a} and \textbf{b} in Figure~\ref{fig:thry:SharpCut}, with the selection function in Equation~\ref{eqn:thry:estComb} derived from it as panel~\textbf{e}. That said, the estimator is:
\begin{widetext}
\begin{align}
	\ln(\mathcal{L}) & = \sum_{\mathrm{galaxies}}  \vphantom{\int} \ln\left( S(F_{\mathrm{sel}}, F_0, \vec{x}) 
		\vphantom{-}\Lsed(F_{\mathrm{sel}}, F_0 | L_0) 
		\Phi(L_0, z)\right)  \nonumber \\
	& \hphantom{=}  - \int 
		S(F_{\mathrm{sel}}, F_0, \vec{x}) 
		\vphantom{-}\Lsed(F_{\mathrm{sel}}, F_0 | L_0) 
		\Phi(L_0, z)
		\d F_{\mathrm{sel}} \d F_0 \d L_0 \d V_c  , \label{eqn:thry:finalLik}
\end{align}
\end{widetext}
where $F_0$ is the observer frame flux $K$-corrected to $L_0$ using an SED fit to all of the galaxy's available photometry, $L_0$ is the luminosity corresponding to $w_0$, $\vec{x}$ is the position in space of the galaxies, and $F_{\mathrm{sel}}$ is the flux which was used to select redshift survey targets, assuming it is not the same as $F_0$. While the selection function, $S(F_{\mathrm{sel}}, F_0, \vec{x})$, can be smooth, it is useful to speed up the calculations by approximating it as flat where it is non-zero, uniform on the sky within the boundary of the survey, and confined to any appropriate redshift limits (that is, a sharp cut). The individual SED fit is what justifies the treatment of $F_0$ and $L_0$ as statistically independent quantities, because otherwise $L_0$ would not be an observable for most galaxies and would have to be integrated over in every term individually, numerically, which would impose a prohibitive performance penalty on the calculation.

The explicit form of \Lsed\ can be derived from Equation~\ref{eqn:thry:mLL}:
\begin{widetext}
\begin{align}
	\vphantom{-}\Lsed(F_{\mathrm{sel}}, F_0 | L_0) & = 
		\frac{\e^{\tau_1 + \tau_2}}{\sqrt{(2\pi)^2 \operatorname{det}(\sigma)}} \cdot \left(\frac{4\pi D_L(z)^2}{(1+z)L_0}\right)^2  \cdot \exp\left(-\frac{1}{2} \sum_{i, j=1}^2 \left[\ell_i - \mu_i\right] [\sigma^{-1}]_{ij} 
		\cdot \left[\ell_j - \mu_j\right] \right) ,\ \mathrm{and} \label{eqn:thry:finalLsed} \\
	\ell_i & = \frac{F_i 4 \pi D_L(z)^2 \e^{\tau_i}}{(1+z) L_0}, \nonumber
\end{align}
\end{widetext}
where $\tau_i$ is the optical depth of foreground dust and gas extinction toward the target galaxy. When numerically computing \Lsed\ it is important that the matrix $\sigma$ in Equation~\ref{eqn:thry:finalLsed} is invertable. A simple model for the noise to signal ratio in the observed fluxes, $F_i$, added to the diagonal entries of $\Sigma_{ij}$ suffices for this purpose (no Einstein summation is assumed):
\begin{align}
	\sigma_{ij} & = \Sigma_{ij} + \left( \delta_{ij} A_i F_i^{B_i} + \sigma_{\tau\,i} \sigma_{\tau\,j}\right) \mu_i \mu_j, \label{eqn:thry:noiseMod}
\end{align}
where $A$ and $B$ are noise model parameters derived from an ordinary least squares fit of $\ln(\sigma_F^2 F^{-2})$ to $\ln(F)$, that is a log-space fit of the squared noise-to-signal ratio to the log of the observed flux. While more realistic models incorporate multiple sources of noise, like the background or positional jitter, a single power law is often sufficient since the color covariance, $\Sigma_{ij}$, is the dominant contribution to $\sigma_{ij}$ at most redshifts. $\sigma_{\tau\,i}$ is a way of including the impact of differences in foreground dust and gas obscuration over the survey field, if the survey was targeted on fluxes that were not extinction corrected; specifically, it is the standard deviation of the dust optical depth of the Milky Way for targets in the survey if selection was based on pre-extinction corrected fluxes. The form of the $\sigma_{\tau\,i}$ contribution to the matrix is set by the fact that each direction has a single $\operatorname{E}(B-V)$ that each $\tau_i$ is proportional to in the extinction model of \cite{Cardelli:1989}, so the uncertainties at different wavelengths are correlated.

The effective selection function in redshift-luminosity space, the $S(L, z)$ from Section~\ref{sec:msr:density}, is directly derivable from \Lsed\ as:
\begin{align}
	S(L, z) & = \int S(F_{\mathrm{sel}}, F_0, \vec{x})\ \Lsed(F_{\mathrm{sel}}, F_0 | L) \d F_{\mathrm{sel}} \d F_0. \label{eqn:thry:selfun}
\end{align}
It is to avoid evaluating the integrals in Equation~\ref{eqn:thry:selfun} numerically for every different galaxy in a survey that Equation~\ref{eqn:thry:estComb} is not recommended as the best LF likelihood estimator. 

It is also arguable that rest frame luminosities, $L_0$, are not actually directly observable outside of a narrow range of redshifts. In that light, the expression of the LF likelihood is:
\begin{align}
	\ln(\mathcal{L}) & = \sum_i \ln( \rho_n(\vec{F}_i, z_i) ) - \int \rho_n(\vec{F}, z) \d^M F \d V \nonumber \\
	\rho_n(\vec{F}, z) & \equiv S(\vec{F}, z) \int \Lsed(\vec{F}| L, z)\ \Phi(L, z) \d L.
\end{align}
This has all of the computational load drawbacks of Equation~\ref{eqn:thry:selfun}. It also has the problem that the normalizing integral, the integral over $\d^M F$, is an integral in a high dimensional space over a finite box, and that further slows numerical performance, even for a Gaussian function.

\subsection{Binned Luminosity Function Estimators}\label{sec:msr:bin}
It is standard practice in most measurements of the LF to include at least one binned estimator to provide an estimate of the LF that has no parametric assumptions going into it. Examining some binned estimators in the presence of the varying selection functions defined in previous sections becomes a necessary step in getting meaningful binned estimates to compare with the parametric estimates in LW17III.
 
The three most popular binned estimators for the LF are: $1/V_{\mathrm{max}}$ \citep{Schmidt:1968}, step-wise maximum likelihood (SWML) \citep{SWML}, and $C^-$ \citep{Lynden-Bell:1971}. $C^-$ is built around estimating the cumulative luminosity function, $\int_L^\infty \Phi(s, z) \d s / \int_{L_0}^\infty \Phi(s, z)\d s$, and does not recover the normalization. SWML is based on the STY version of the likelihood, and ends up with correlated errors between the bins from the technique used to recover $\Phi$'s normalization. The $1/V_{\mathrm{max}}$ estimator has very well explored drawbacks, sensitivity to clustering \citep{Takeuchi:2000}, and trouble near the luminosity limits \citep{Page:2000}.

\cite{Miyaji:2001} proposed an estimator they named $N^{\mathrm{obs}} / N^{\mathrm{mdl}}$:
\begin{align}
	\Phi(L_i, z_i) & = \Phi^{\mathrm{mdl}}(L_i, z_i) \cdot \frac{N^{\mathrm{obs}}_i}{N^{\mathrm{mdl}}_i},
\end{align}
where $\Phi(L_i, z_i)$ is the value of the LF in $L$-$z$ bin $i$, ``$\mathrm{mdl}$" is short for ``model," $\Phi^{\mathrm{mdl}}$ is an approximate LF model, $N^{\mathrm{mdl}}_i$ is the expected number of observations in bin $i$ according to approximate model LF $\Phi^{\mathrm{mdl}}$, and $N^{\mathrm{obs}}_i$ is the number of objects observed to be in bin $i$. \cite{Miyaji:2001} did not include a derivation of this estimator, so one follows here, including some refinements.

The definition of the luminosity function implies that the mean number of galaxies in a bin is:
\begin{align}
	\langle N_i \rangle & = \int_{\mathrm{bin}} S(L, z) \Phi(L, z) \d L \d V.
\end{align}
The weighted mean value theorem implies that:
\begin{align}
	\Phi(L', z') & = \frac{\langle N_i \rangle}{\int_{\mathrm{bin}} S(L, z) \d L \d V}, \label{eqn:thry:bin0}
\end{align}
for some $(L', z')$ in the $L$-$z$ bin where the selection function, $S$, is non-zero. The maximum likelihood estimator for $\langle N_i \rangle$ is the observed number, $N_i$ (if the $N_i$ are Poisson distributed), thus:
\begin{align}
	\Phi(L', z') & = \frac{ N_i }{\int_{\mathrm{bin}} S(L, z) \d L \d V}. \label{eqn:thry:bin1}
\end{align}

In the approximation that $(L', z')$ is at the center of the box and $S$ is a step function, Equation~\ref{eqn:thry:bin1} is the estimator from \cite{Page:2000}, which corrects for the near luminosity limits problems of the $1/V_{\mathrm{max}}$ estimator. Recovering the $1/V_{\mathrm{max}}$ estimator requires (in order): making the step function $S$ approximation, making the bins infinitesimally thin in the luminosity direction, and finally binning infinitesimal luminosity bins back down into the coarse bins desired, yielding (for log-space binning):
\begin{align}
	\Phi(L_i, z_i) & = \frac{1}{S(L_i, z_i) L_i \Delta \ln L_i} \sum_j \frac{1}{\Delta V_j},
\end{align}
where $j$ runs over objects in bin $i$, $\Delta V_j \equiv V(z_{\mathrm{hi}}) - V(z_{\mathrm{lo}})$, $z_{\mathrm{hi}}$ is the smaller of the upper redshift of the bin and $z_{\mathrm{max}}$, the maximum redshift at which the galaxy is observable, and similar in the opposite sense for $z_{\mathrm{lo}}$. $\Delta V_j$ is called $V_a$, for volume available, in \cite{Avni:1980}.

The $N^{\mathrm{obs}} / N^{\mathrm{mdl}}$ estimator of \cite{Miyaji:2001} comes from using an approximate model for the LF, $\Phi^{\mathrm{mdl}}$, to correct the estimate from Equation~\ref{eqn:thry:bin1} from being accurate at some unknown point in the bin to being an estimate of $\Phi$ at the center of the bin. This process is formally identical to how flux corrections for different SED shapes in astronomy are done:
\begin{align}
	\Phi(L_i, z_i) & = \frac{\Phi^{\mathrm{mdl}}(L_i, z_i)}{\Phi^{\mathrm{mdl}}(L',z')}\cdot \frac{ N_i }{\int_{\mathrm{bin}} S(L, z) \d L \d V}, \nonumber \\
	& = \Phi^{\mathrm{mdl}}(L_i, z_i) \cdot \frac{N_i}{\langle N^{\mathrm{mdl}}_i\rangle },\label{eqn:thry:bin}
\end{align}
where the second line undoes the weighted mean value theorem. 

The downside of the $N^{\mathrm{obs}} / N^{\mathrm{mdl}}$ estimator, which can also be described as a corrected Poisson estimate, is that there is some dependence on the model LF introduced by the correction. A detailed exploration of this sensitivity is beyond the scope of this work, but any model closer to the true LF than $\Phi^{\mathrm{mdl}} = \mathrm{constant}$, the model implicitly assumed when approximating $L'$ and $z'$ as being at the center of the bin in Equation~\ref{eqn:thry:bin1}, will give improved results.

\subsection{Error Analysis}\label{sec:msr:err}
The uncertainties in the binned estimators can be estimated using propagation of errors and the assumption that Poisson/shot noise describes all the variance:
\begin{align}
	\sigma_{1/V_{\mathrm{max}},\, i}^2 & = \left(\frac{1}{L_i \Delta \ln L_i}\right)^2 \sum_j \frac{1}{(S_j  \Delta V_j)^2},\ \mathrm{and} \nonumber \\
	\sigma_{N^{\mathrm{obs}} / N^{\mathrm{mdl}},\, i}^2 & = \left(\frac{\Phi^{\mathrm{mdl}}_i}{\langle N^{\mathrm{mdl}}_i\rangle}\right)^2 N_i. \label{eqn:thry:poisErr}
\end{align}
The formula for $\sigma_{1/V_{\mathrm{max,i}}}^2$ comes from doing propagation of errors in the process of binning down the estimate in the luminosity direction. Equations~\ref{eqn:thry:poisErr} underestimate the error because they do not account for a number of factors, including: cosmic variance, redshift uncertainties, or luminosity uncertainties (including the contribution from the variety of SED colors discussed in \cite{Lake:2016}. How to handle these factors in the presence of smoothly varying selection functions is beyond the scope of this work.

\section{Luminosity Function Parametrizations}\label{sec:Parameterization}
\begin{figure*}[htb]
	\begin{center}
	\includegraphics[width=\textwidth]{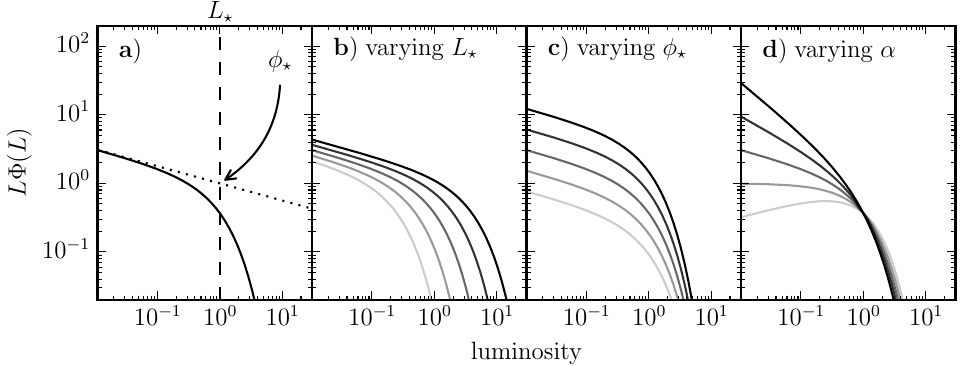}
	\end{center}
	\caption{ Illustrations of the shape of the Schechter function (times $L$) in log-log space, and how that shape changes as the parameters, $L_\star$, $\phi_\star$, and $\alpha$, vary. Panel~\textbf{a} is a labeled illustration with $L_\star = 1$, $\phi_\star=1$, and $\alpha=-1.25$. The vertical dashed line shows $L_\star$, and the dotted line shows the faint end slope (asymptotic power law for faint luminosity, log-log slope in this graph is $\alpha + 1$) and how $\phi_\star$ is the $y$ value of where the dashed and dotted lines intersect. Panels~\textbf{b}--\textbf{d} show how the Schechter function changes with $L_\star$, $\phi_\star$, and $\alpha$, respectively. The lines are increasingly dark with increasing magnitude of the parameter being varied. }
	\label{fig:thry:Schech}
\end{figure*}

For galaxies, the standard practice is to model the LF as a Schechter function:
\begin{align}
	\Phi(L, z) & = \frac{\phi_\star(z)}{L_\star(z)}\left(\frac{L}{L_\star(z)}\right)^\alpha \mathrm{e}^{-L/L_\star(z)}. \label{thry:eqn:Schech}
\end{align}
This form for the LF was derived in \cite{Schechter:1976} under the assumption that galaxies build up in luminosity through a process of merging. While mergers are not the only way for galaxies to build up in luminosity, the function still works well empirically. Figure~\ref{fig:thry:Schech} contains an illustration of the shape the Schechter function takes, in log-log space, including labels for $L_\star$ and $\phi_\star$, and illustrations of how the shape changes when each of the parameters changes, as they would for evolution in time.

In principle all three of the parameters in $\Phi$ could evolve with $z$, but measuring the evolution in $\alpha$  requires data that is deep over a broad range of redshifts, so most works assume it is constant. For $L_\star$ and $\phi_\star$ the most frequently used parameterizations are those of \cite{Lin:1999} ($L_\star \propto 10^{0.4 Q z}$ and $\phi_\star \propto 10^{0.4 P z}$, with $P$ and $Q$ constants). While these parameterizations work fine in a small range of redshifts, they allow for unphysically unbounded evolution at high $z$. This work, therefore, introduces the following similar parameterizations in terms of $t_L(z)$, the lookback time at redshift $z$:
\begin{align}
	\phi_\star & = \phi_0 \mathrm{e}^{- R_\phi t_L(z) },\ \mathrm{and} \nonumber \\
	L_\star & = L_0 \mathrm{e}^{- R_L t_L(z) } \left(1 - \frac{t_L(z)}{t_0}\right)^{n_0}, \label{thry:eqn:Param}
\end{align}
where $R_\phi$ is the density evolution rate, $R_L$ is the luminosity evolution rate, $n_0$ is the initial luminosity index, and $t_0$ is the lookback time at which galaxies first lit up (some redshift between reionization and recombination). Given the definitions of the parameters, $L_\star(t_L = t_0) = 0$ is a boundary condition, requiring that $n_0 > 0$. Parametrizing $\phi_\star$ evolution in terms of lookback time also guarantees that it will remain finite for all redshifts regardless of whether $R_\phi$ is positive or negative. The addition of a power law to the evolution of $L_\star$ is inspired by the model of star formation rate (SFR) in individual galaxies used in \cite{Lee:2010} and tested against simulations in \cite{Simha:2014}: $\operatorname{SFR}(t) \propto t \exp(-t / \tau)$, where $t$ is measured from the onset of star formation. 


One advantage of the form for $L_\star$ in Equation~\ref{thry:eqn:Param} is that the model now contains an estimate of where the big galaxies hit their peak luminosity, the redshift at which $L_\star$ has a maximum, $z_\star$:
\begin{align}
	t_L(z_\star) & = t_0 + \frac{n_0}{R_L}. \label{thry:eqn:zstar}
\end{align}
Note that $L_\star$ only has a maximum at finite time if $R_L<0$; when this condition is not met $L_\star$ increases without bound for all future time, putting the maximum at $z=-1$.

\begin{figure}[htb]
	\begin{center}
	\includegraphics[width=.48\textwidth]{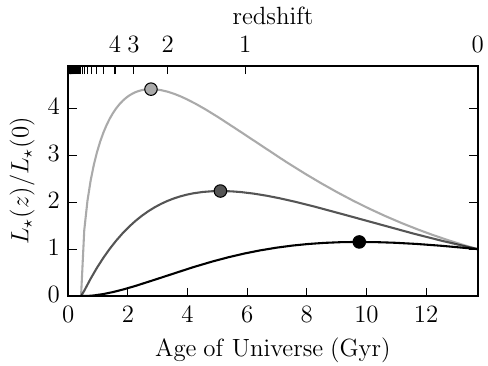}
	\end{center}
	\caption{ Illustration of how $L_\star$ evolves over cosmic time in the model introduced in this work. The value of $R_L$ used is $-3 t_H^{-1}$, the value of $t_0$ is the lookback time of the redshift of recombination ($z=1088.16$), and the values of $n_0$ are $0.5$, $1$, and $2$, in increasing darkness. The circular dots are placed at the time of peak $L_\star$ for each curve. The cosmology assumed is based on the WMAP 9 year $\Lambda$CDM cosmology \citep{Hinshaw:2013}. }
	\label{fig:thry:LstarEvo}
\end{figure}

A graph of how our simple model for $L_\star(z)$ evolves with cosmological time can be found in Figure~\ref{fig:thry:LstarEvo}. The cosmology assumed for the figure is based on the WMAP 9 year $\Lambda$CDM cosmology \citep{Hinshaw:2013}\footnote{\url{http://lambda.gsfc.nasa.gov/product/map/dr5/params/lcdm\_wmap9.cfm}}, with flatness imposed, yielding:  $\Omega_M = 0.2793,\ \Omega_\Lambda = 1 - \Omega_M$, and $H_0 = 70 \km \second^{-1} \Mpc^{-1}$ (giving Hubble time $t_H = H_0^{-1} = 13.97\operatorname{Gyr}$).

\subsection{Derived LF Parameters}
A number of physical quantities are derivable from the luminosity function. The first pair of such quantities are the density of galaxies brighter than some cutoff, the zeroth moment of the Luminosity function, and its specific rate of change:
\begin{align}
	n_g(L > L_{\mathrm{min}}) & = \int_{L_{\mathrm{min}}}^\infty \Phi(L, z) \operatorname{d}L, \ \mathrm{and}\\
	R_n & \equiv - \frac{\partial \ln(n_g)}{\partial t_L} \label{eqn:thry:specNDrate},
\end{align}
respectively. But for the low luminosity cutoff, necessary to make $n_g$ finite, both quantities would be universal; that is, independent of the bandpass they are measured in. Because $R_n$ depends only weakly on the parameter $L_{\mathrm{min}} / L_\star$ (small for most redshifts), it is approximately universal, making it extremely useful to compare measurements of LF evolution regardless of which luminosity was used in the measurement.

For a general Schechter function $n_g$ and $R_n$ are:
\begin{align}
	n_g & = \phi_\star \Gamma\left(\alpha + 1, \frac{L_{\mathrm{min}}}{L_\star}\right),\ \mathrm{and}\\
	R_n & = -\phi_\star^{-1} \frac{\partial \phi_\star}{\partial t_L} - \left(\frac{\left[\frac{L_{\mathrm{min}}}{L_\star}\right]^{\alpha + 1} \e^{-L_{\mathrm{min}} / L_\star}}{\Gamma\left(\alpha + 1,\, \frac{L_{\mathrm{min}}}{L_\star}\right)}\right) L_\star^{-1}\frac{\partial L_\star}{\partial t_L} \nonumber \\
	& \hphantom{=}\  - \left(\frac{\int_{L_{\mathrm{min}} / L_\star}^\infty x^\alpha \ln(x)\, \e^{-x} \d x }{\Gamma\left(\alpha + 1,\, \frac{L_{\mathrm{min}}}{L_\star}\right)}\right) \frac{\partial \alpha }{\partial t_L} \nonumber \\
	& \approx -\phi_\star^{-1} \frac{\partial \phi_\star}{\partial t_L} + \operatorname{min}\left(\alpha + 1,\, 0\right) L_\star^{-1} \frac{\partial L_\star}{\partial t_L} \nonumber \\
	& \hphantom{\approx}\ - \left(\left[\ln\left(\frac{L_{\mathrm{min}}}{L_\star}\right) - \frac{1}{\alpha + 1}\right] \Theta(-\alpha-1)  \right. \nonumber \\
	& \hphantom{\approx - (} \left. \vphantom{\frac{L_{\mathrm{min}}}{L_\star}} + \psi^{(0)}(\alpha + 1) \Theta(\alpha + 1) \right) \frac{\partial \alpha }{\partial t_L} , \label{eqn:thry:RnEqns}
\end{align}
where $\Gamma(a, z)$ is the (upper) incomplete gamma function, $\psi^{(0)}(x)$ is the digamma function, and $\Theta(x)$ is the unit step function (Heaviside). As long as $\alpha$ is (nearly) constant, and $L_{\mathrm{min}} \ll L_\star$, the band dependence of $R_n$ is vanishingly small, making $R_n$ universal at most times. 
The time when the universality of $R_n$ fails is at early times (high redshifts near reionization) when $L_\star$ is near $L_{\mathrm{min}}$. For the parametrization defined here, and the one defined in \cite{Lin:1999}, respectively:
\begin{align}
	R_n & = R_\phi - \operatorname{min}(1+\alpha, 0) \left[ R_L + \frac{n_0}{t_0 - t_L} \right],\ \mathrm{and} \label{eqn:thry:myRn} \\
	 R_n & = - \frac{2}{5}\ln(10) \left(P - \operatorname{min}(1+\alpha, 0)Q\right) \frac{\operatorname{d}z}{\operatorname{d}t_L}. 
\end{align}


For the purposes of fitting a LF to data in a Bayesian framework, it can be useful to make the substitution $R_n(0) = R_\phi - (1+\alpha) (R_L + n_0 t_0^{-1})$ to replace $R_\phi$. This substitution is profitable because we have observed $R_n(0)$ to be less correlated with $\alpha$ and $L_\star$ than $R_\phi$ with real data sets.

The CLD, $\rho_L$, and its specific rate of change are defined as:
\begin{align}
	\rho_L & = \int_0^\infty L \Phi(L, z) \operatorname{d}L,\ \mathrm{and}\\
	R_\rho & = - \frac{\partial \ln( \rho_L )}{\partial t_L}.
\end{align}
Applying the definitions to the Schechter function parametrization of $\Phi$ yields:
\begin{align}
	\rho_L & = \phi_\star L_\star \Gamma(\alpha + 2),\ \mathrm{and} \label{thry:eqn:LD}\\
	R_\rho & = -\phi_\star^{-1} \frac{\partial \phi_\star}{\partial t_L} -
		L_\star^{-1} \frac{\partial L_\star}{\partial t_L} - \psi^{(0)}(\alpha + 2) \frac{\partial \alpha}{\partial t_L}, \label{thry:eqn:LDrate}
\end{align}
where $\psi^{(0)}(x)$ is the digamma function. When further specialized to the parametrization defined here and the one from \cite{Lin:1999}, Equation \ref{thry:eqn:LDrate} gives:
\begin{align}
	R_\rho & = R_\phi + R_L + \frac{n_0}{t_0 - t_L},\ \mathrm{and}  \\
	R_\rho & = -\frac{2}{5} \ln(10) (P + Q) \frac{\operatorname{d} z}{\operatorname{d} t_L},
\end{align}
respectively. The redshift at which the CLD hits a maximum, $z_\rho$, can also be found for the parametrization defined here by solving for $z_\rho$ in the equation:
\begin{align}
	t_L(z_\rho) & = t_0 + \frac{n_0}{R_\phi + R_L}. \label{thry:eqn:zj}
\end{align}
Like with the Equation~\ref{thry:eqn:zstar}, existence of a solution at finite time to Equation~\ref{thry:eqn:zj} requires the assumption that the present CLD is decreasing, $R_\phi + R_L < 0$, with a maximum at $z=-1$ otherwise.

The third quantity of interest is closely related to to the flux counts density on the sky, $\operatorname{d}N / (\operatorname{d} F \operatorname{d}\Omega)$. It is a well known result that for an infinite Euclidean universe with a static LF that $\operatorname{d}N / (\operatorname{d} F \operatorname{d}\Omega) \propto F^{-5/2}$. It is possible to show that the constant of proportionality, here called $\kappa_\star$, can be calculated from the Schechter LF parameters:
\begin{align}
	\kappa_\star & = \frac{\phi_\star L_\star^{3/2}}{\Omega_{\mathrm{sky}} 4\pi^{1/2}} \Gamma\left(\alpha + \frac{5}{2}\right), \label{eqn:thry:kappastar}
\end{align}
as long as $L_\star$, $\alpha$, and $\phi_\star$ are assumed constant.

The utility of parametrizing the LF in terms of $\kappa_\star$ is that, as the LF measurements in LW17III shows, real data constrains it more than $\phi_\star$, making it less correlated to the other parameters in the Schechter function. Further, when performing a Markov Chain Monte Carlo analysis of a model, it is important to have reasonable bounds for the parameters in the data to constrain the search space. For $\kappa_\star$, it is possible to produce a histogram of fluxes in log-log axes, weighted to remove the Euclidean power law slope ($-5/2$) and the effects of redshift, and read off a reasonable range of values that $\kappa_\star$ can take. To remove the non-evolution effects of redshift it is sufficient to histogram the flux the source would have if the universe were static and Euclidean, $\tilde{F} \equiv L / (4\pi D_{\mathrm{cT}}^2(z))$, where $D_{\mathrm{cT}}(z)$ is the comoving distance transverse to the line of sight (also called $D_M$). $L_\star$ can be similarly constrained by visual inspection of luminosity histograms. The fundamental definition of $\phi_\star$, $\Phi(L_\star) \propto \phi_\star L_\star^{-1}$, means that the value of $\phi_\star$ is contingent on the value of other observables, and therefore less straightforward to read off a reasonable range from inspection of a graph. 

\begin{figure}[htb]
	\begin{center}
	\includegraphics[width=0.48\textwidth]{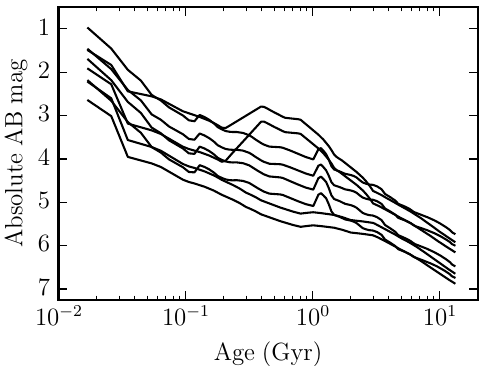}
	\end{center}
	
	\caption{ Absolute magnitude evolution tracks of simple stellar populations using the models from \cite{Bruzual:2003} and \cite{Maraston:2005}, in $K_s$ and $\mathrm{W1}$ bands, and with Salpeter and Chabrier IMFs. The different combinations are unlabeled because what is important is that the luminosity decays in a power law like fashion (approximately $\propto t^{-1/2}$), and not an exponential one. }
	\label{fig:thry:SSP}
\end{figure}

The models produced here for the evolution of $L_\star$ and $\phi_\star$ have the advantage that they are computationally simple. While their form is based on improving the early cosmic time behavior of the models compared to the models of \cite{Lin:1999}, there is much room for improvement yet. The models in this work imply, for instance, that the long time CLD will exponentially decay to zero. This behavior conflicts with a lower bound on the rate at which the CLD can decrease implied by at least two stellar population synthesis models. Assuming that the CLD is already dominated by stars and not AGN, the fastest the CLD can decay is the rate it would drop if all star formation were to cease abruptly. Thus the CLD decay rate is bounded from below by the luminosity decay rate of a simple stellar population (SSP). Figure~\ref{fig:thry:SSP} shows the evolution of an SSP's absolute magnitude in two infrared bands that bracket $\lambda = 2.4\micron$ (2MASS $K_s$ and \WISE\ W1) with two different initial mass functions (Salpeter and Chabrier) from two different population synthesis models \citep{Bruzual:2003, Maraston:2005} as implemented by the EzGal software package \citep[described in:][]{Mancone:2012}. The different combinations are not labeled because the details are not important here, only the overall trend that they all decay in luminosity like a power law (approximately $\propto t^{-1/2}$), which is much slower than an exponential decay.

\section{Discussion} \label{sec:discussion}
We summarize the commonly adopted luminosity function estimate processes, including the estimators from: \cite{Schmidt:1968}, \cite{STY}, \cite{Marshall:1983}, \cite{Page:2000}, and \cite{Miyaji:2001}. This work also introduces improvements to the luminosity function measuring process from the point of view of mathematical rigor, compatibility with the physical constraints implied by the current understanding of cosmology, and raw statistical performance (statistical orthogonality of parameters). While that last property is not demonstrated here, it will be apparent when the techniques defined here are applied to the real world data sets of \citep[LW17II]{Lake:2017b} in LW17III. In particular, we show that the cuts to the data sets necessary to render the selection functions approximately flat remove 27--81\% of the data in the set, depending on how flat the selection function is required to be. Making such a cut is not sufficient to relieve the computational burden required to address the selection biases addressed in this work, however, because determining the shape of the flat region requires roughly the same effort on the part of the scientist analyzing the data.

We hope that the decision to present not just the final form of the estimator, but also its derivation, will make it easier for future users of these techniques to adapt them to their specific needs, and even to make improvements as computational techniques and hardware improve. Areas where there is significant room for improvement in the model include:
\begin{itemize}
	\item the model does not yet make any allowance for the effect of clustering/cosmic variance,
	\item the model does not make allowance for the effects of finite galaxy size in relation to mergers,
	\item the model does not account for line of sight overlap, coincidental or otherwise (confusion and collision),
	\item the long time evolution of $L_\star$ and $\phi_\star$ defined here conflicts with the behavior of population synthesis models, and
	\item the Gaussian approximation for \Lsed\ restricts its applicability to a limited range of wavelengths where the standard deviation of the SED is less than its mean.
\end{itemize}

\bibliography{LFtheoryBib}

\end{document}